\documentclass[twoside]{IEEEtran}
\usepackage{cite}

   \usepackage[pdftex]{graphicx}
\usepackage{amsmath}

\usepackage{algorithmic}

\usepackage{array}
\usepackage{fixltx2e}
\usepackage{url}

\usepackage{xcolor,soul,framed} 
\colorlet{shadecolor}{yellow}
\usepackage[pdftex]{graphicx}
\graphicspath{{../pdf/}{../jpeg/}}
\DeclareGraphicsExtensions{.pdf,.jpeg,.png}
\usepackage{algorithm}
\usepackage{amsfonts}
\usepackage{mathtools}
\usepackage{afterpage}

\providecommand{\keywords}[1]{\textbf{\textit{Index Terms---}} \textbf{\small #1}}
\usepackage{fancyhdr}
\usepackage{afterpage}
\usepackage{scalerel}
\usepackage{tikz}
\usetikzlibrary{svg.path}

\definecolor{orcidlogocol}{HTML}{A6CE39}
\tikzset{
	orcidlogo/.pic={
		\fill[orcidlogocol] svg{M256,128c0,70.7-57.3,128-128,128C57.3,256,0,198.7,0,128C0,57.3,57.3,0,128,0C198.7,0,256,57.3,256,128z};
		\fill[white] svg{M86.3,186.2H70.9V79.1h15.4v48.4V186.2z}
		svg{M108.9,79.1h41.6c39.6,0,57,28.3,57,53.6c0,27.5-21.5,53.6-56.8,53.6h-41.8V79.1z M124.3,172.4h24.5c34.9,0,42.9-26.5,42.9-39.7c0-21.5-13.7-39.7-43.7-39.7h-23.7V172.4z}
		svg{M88.7,56.8c0,5.5-4.5,10.1-10.1,10.1c-5.6,0-10.1-4.6-10.1-10.1c0-5.6,4.5-10.1,10.1-10.1C84.2,46.7,88.7,51.3,88.7,56.8z};
	}
}

\newcommand\orcidicon[1]{\href{https://orcid.org/#1}{\mbox{\scalerel*{
				\begin{tikzpicture}[yscale=-1,transform shape]
				\pic{orcidlogo};
				\end{tikzpicture}
			}{|}}}}

\usepackage{hyperref} 
\usepackage[]{footmisc}
\begin{document}
%
\title{Reconfigurable Intelligent Surface-Empowered MIMO Systems}
%
%
%

\author{Aymen~Khaleel $^{\orcidicon{0000-0001-5258-4720}}$,~\IEEEmembership{Student Member,~IEEE}
        and Ertugrul~Basar $^{\orcidicon{0000-0001-5566-2392}}$,~\IEEEmembership{Senior Member,~IEEE} 
\thanks{\hspace{-0.3cm}Manuscript received April 2, 2020; revised June 7, 2020 and July 20, 2020; accepted July 22, 2020. Date of current version July 22, 2020. This work was supported by the Scientific and Technological Research Council of Turkey (TUBITAK) under Grant 117E869. \textit{(Corresponding author: Ertugrul Basar.)}
	
	The authors are with the Communications Research and Innovation Laboratory (CoreLab),  Department of Electrical and Electronics Engineering, Ko\c{c} University, Sariyer 34450, Istanbul, Turkey. \mbox{Email: akhaleel18@ku.edu.tr, ebasar@ku.edu.tr}.
}
}
\maketitle
\thispagestyle{fancy}
\cfoot{{\scriptsize 1937-9234 © 2020 IEEE. Personal use is permitted, but republication/redistribution requires IEEE permission.\\See https://www.ieee.org/publications/rights/index.html for more information.\\}}
 %
\begin{abstract}
Reconfigurable intelligent surface (RIS)-assisted communications appear as a promising candidate for future wireless systems due to its attractive advantages in terms of implementation cost and end-to-end system performance. In this paper, two new multiple-input multiple-output (MIMO) system designs using RISs are presented to enhance the performance and boost the spectral efficiency of state-of-the-art MIMO  communication systems. Vertical Bell Labs layered space-time (VBLAST) and Alamouti's schemes have been considered in this study and RIS-based simple transceiver architectures are proposed. For the VBLAST-based new system, an RIS is used to enhance the performance of the nulling and canceling-based sub-optimal detection procedure as well as to noticeably boost the spectral efficiency by conveying extra bits through the adjustment of the phases of the RIS elements. In addition, RIS elements have been utilized in order to redesign  Alamouti's scheme with a single radio frequency (RF) signal generator at the transmitter side and to enhance its bit error rate (BER) performance. Monte Carlo simulations are provided to show the effectiveness of our system designs and it has been shown that they outperform the reference schemes in terms of BER performance and spectral efficiency.
\end{abstract}

\keywords{\hspace{-0.3cm}Reconfigurable intelligent surface (RIS), MIMO systems, error probability analysis, VBLAST, Alamouti's scheme.}


%
\IEEEpeerreviewmaketitle

\section{Introduction}
\IEEEPARstart{R}{econfigurable} intelligent surfaces (RISs) have received significant attention from the wireless communication community as effective, cheap, reconfigurable, easy to deploy, and passive system modules that can be used to control the wireless propagation environment by re-engineering the electromagnetic waves \cite{Wireless_communications_through}, \cite{TowardsAided}. Manipulating the propagation environment using RISs has been regarded as a promising candidate for the next-generation wireless technologies such as Terahertz communications, non-orthogonal multiple access (NOMA), and low-cost massive multiple-input multiple-output (MIMO) systems. Without loss of generality, mitigating the fading channel impairments and compensating for the propagation losses in order to enhance the signal quality at the receiver side are the main objectives behind the development of this technology. Nevertheless, RISs can also be utilized to minimize the transmitted signal power, to boost the system transmission capacity, and to enhance the physical layer security \cite{RIS-Survey}, \cite{RIS-Survey2}.  

In the preliminary study of \cite{Transmission_conference}, RISs are employed for two different purposes. First, an RIS is used to realize an ultra-reliable communication scheme that operates at considerably low signal-to-noise ratio (SNR) values. While in the second scenario, an RIS is used as an access point to create virtual phase shift keying (PSK) symbols at the receiver side. The latter concept is also used to perform index modulation at the receiver side \cite{surface-based}. Considering a dual-hop communication scenario, the authors in \cite{SSK} proposed an RIS-based space shift keying system where the RIS is used as a reflector which is positioned between a transmitter with multiple antennas and a receiver with a single antenna. In \cite{Increasing_indoor_spectrum_sharing}, the indoor multiple-user network-sharing capacity is enhanced by optimally adjusting the phases of a passive reconfigurable reflect-array to cancel the interference and enhance the users' signal quality. An energy efficient multiple-input single output (MISO) system is proposed in \cite{Energy_Efficient_Multi-User}, by jointly optimizing the transmit powers of the users and the phases of RIS elements. In \cite{SWPT_RIS}, the authors proposed an RIS-assisted simultaneous wireless information and power transfer system, where an information decoding set and energy harvesting set of single-antenna  receivers are served by a multiple-antenna access point. In \cite{AN_RIS}, the authors investigated the use of artificial noise in order to increase the secrecy rate in an RIS-based communication system, where a single-antenna user is served by a multiple-antenna transmitter in the presence of multi-antenna eavesdroppers. In \cite{Joint_Active_and_Passive_Beamforming} and \cite{Joint_Active2}, the received signal power for a MISO user is maximized by optimizing the active beamforming at the transmitter jointly with the passive beamforming at the RIS by adjusting its phase shifters. The latter concept is also used in \cite{MISO_wireless}, where the beamformer at the access point and RIS are jointly optimized in order to increase the spectral efficiency for an  RIS-assisted multi-user MISO system. In \cite{Asymptotic_analysis}, an RIS-assisted multi-user MISO system is considered with different channel types where RIS phase optimization is utilized in order to maximize the minimum signal-to-noise-and-interference ratio (SNIR). RIS is used in \cite{RIS-RANK} to improve the channel rank for MIMO systems by adding additional multipaths with distinctively different spatial angles in addition to the low-rank direct channel path. RIS reflection coefficients and the transmit covariance matrix are jointly optimized in \cite{RIS-CAPACITY} in order to maximize the capacity of a point-to-point MIMO system. In \cite{RIS-ESTIMATION}, the authors consider the channel estimation problem in multi-user MIMO systems and propose an uplink channel estimation protocol to estimate the cascaded channel from the base station to the RIS and from the RIS to the user. The use of passive intelligent mirrors (PIM) with a multi-user MISO downlink system is investigated in \cite{Sum-rate}, where the transmit powers and the PIM reflection coefficients are designed to maximize the sum-rate considering the individual quality of service for mobile users. An overview of the holographic MIMO surface is presented in \cite{Holographic}, where the authors investigated their hardware architectures, functionalities, and characteristics. In \cite{Deep}, the authors exploited the deep learning reinforcement to jointly obtain the optimum beamforming matrix and the RIS phase shifts where the introduced algorithm learns directly from the environment and updates the beamforming matrix and RIS phase shifts accordingly. Based on cosine similarity theorem, a low-complexity RIS phase shift design algorithm is proposed in \cite{low_comx}, where the RIS is used to assist a MIMO communication system. In \cite{Sim_RIS1} and \cite{Sim_RIS2}, the authors investigated physical channel modeling for mmWave bands considering indoor and outdoor environments. Furthermore, the authors provided an open-source comprehensive channel simulator  which can be used to examine the different channel models discussed in their work. However, the use of RISs to boost the spectral efficiency and/or reliability of existing MIMO systems along with applications of index modulation (IM) is not well explored in the open literature. 
  Against this background, two new RIS-assisted communication schemes are presented in this paper by focusing on the integration of RISs into the existing MIMO systems in a simple and effective way. VBLAST \cite{VBLAST} and Alamouti's schemes \cite{Alamouti} are considered in this study as the most common and practical MIMO schemes while a generalization to other advanced MIMO signaling schemes might be possible using our concept. We summarize the main contributions of this paper as follows:
  \begin{itemize}[
  	\setlength{\IEEElabelindent}{\dimexpr-\labelwidth-\labelsep}
  	\setlength{\itemindent}{\dimexpr\labelwidth+\labelsep}
  	\setlength{\listparindent}{\parindent}
  	]
    \item We propose an RIS-assisted Alamouti's scheme in which we redesign the classical Alamouti's scheme with a single RF signal generator at the transmitter side instead of two RF chains.  
    \item We show that our RIS-assisted Alamouti's scheme preserves the diversity order of the classical Alamouti's scheme and provides a significant BER performance enhancement.
    \item We propose an RIS-assisted and IM-based VBLAST scheme using nulling and cancelling-based sub-optimal detection with zero forcing (ZF) technique. Compared to the classical VBLAST scheme, we show that our proposed RIS-assisted and IM-based VBLAST scheme provides superior performance in terms of the spectral efficiency and the BER performance.
    \item For RIS-assisted and IM-based VBLAST scheme, we propose two novel nulling-based optimal and sub-optimal detectors to detect the indices of the  antennas targeted by the IM.
        \end{itemize} 
    
     In RIS-assisted Alamouti's scheme, at the receiver side, the classical Alamouti's detector is used to recover the transmitted symbols, assuming that the channel state information (CSI) is available at this unit. It is worth noting that, for our RIS-assisted scheme, no CSI is needed at the RIS side, which reduces the overhead for CSI acquisition and simplifies the RIS design. Compared to the blind RIS-AP scheme in \cite{Transmission_conference} and the classical Alamouti's scheme, our results show that the proposed scheme provides a significant improvement in the BER performance. Furthermore, theoretical analysis and simulations results of the RIS-assisted Alamouti's scheme show that the same concept can be generalized to space-time block code (STBC) systems. In other words, in a large scale MIMO setup, our scheme can replace a large number of RF chains at the transmitter side by a single RF signal generator. Furthermore, our RIS-assisted scheme can provide a significant BER performance enhancement while preserving the diversity order of the STBC system.
     \cfoot{}
    In RIS-assisted and IM-based VBLAST scheme, the RIS can be operated in multiple modes with and without IM. With IM, the RIS eliminates the channel phases between a specific transmit-receive antenna pair, which is selected according to the additional information bits in an IM fashion. On the other hand, without IM, the RIS eliminates the channel phases between a fixed and predefined antenna pair in order to provide the maximum BER performance enhancement. In this scheme, the CSI between each transmit-receive antenna pair through the RIS is required at both the RIS and the receiver side. At the receiver side, the IM bits are obtained using our novel nulling-based detectors while the transmitted symbols will be detected as in the plain VBLAST scheme using ZF-based nulling and cancelling algorithm without requiring additional signal processing steps \cite{VBLAST}. The proposed schemes are simple in design and do not require major modifications for the state-of-the-art systems. Furthermore, comprehensive computer simulations are provided in this study under realistic environment setups to assess their practical feasibility.

The rest of the paper is organized as follows. In Section II, we introduce the system model of the RIS-assisted Alamouti's scheme and evaluate its symbol error probability (SEP). Section III introduces the RIS-assisted and IM-based VBLAST scheme, the nulling-based detectors, and the analysis of the computational complexity of the receiver. In Section IV, we provide our computer simulation results and comparisons along with path loss models considered in these simulations. Finally, conclusions are given in Section V.
\section{RIS-Assisted Alamouti's scheme: signal model and error performance analysis}
In the proposed RIS-assisted Alamouti's scheme, an unmodulated carrier signal is being transmitted from a low-cost RF signal generator close to the source (S) unit. The RF signal generator contains an RF digital-to-analog converter with an internal memory and a power amplifier as discussed in \cite{Transmission_conference}. 
      \begin{figure}[t]
	\begin{center}
		\includegraphics[width=9cm]{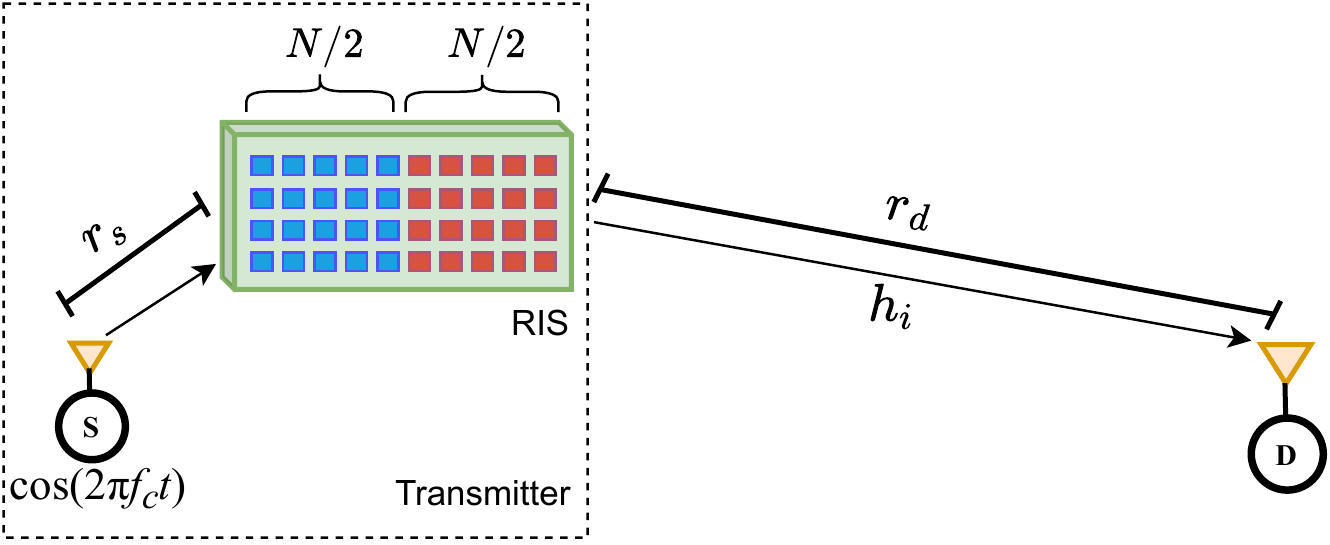}
		\caption{RIS-assisted Alamouti's scheme with the RIS as the transmitter.}\label{Vs RIS-AP}		
	\end{center}
	\vspace{-0.5cm}
\end{figure}
Fig. 1 shows the block diagram of the proposed scheme  where  $r_s$ and $r_d$ are the distances (in meters) of S-RIS and RIS-D, respectively. $r_s$ is selected in a way that the channel between S and RIS is assumed to be line-of-sight (LOS) dominated. In our setup, the RIS is divided into two parts each having $N/2$ elements adjusted to a common reflection phase value. Each part employs two different common reflection phase values over two time slots. The proposed system emulates the Alamouti's scheme by adjusting the phases of the RIS elements to modify the RF carrier signal and invoke the phases of the two data symbols. In this way, the Alamouti's scheme can be redesigned with a single RF signal generator instead of two full RF chains at the transmitter side. The RIS is a blind one with respect to the CSI, while its intelligent stems from the fact that it adjusts the unmodulated RF signal to mimic the PSK symbol phases. In this study, due to the high path loss experienced by the signals reflected from the RIS, the power of the signals reflected from the RIS for two or more times is ignored and only the first reflection is considered in our signal model \cite{Joint_Active_and_Passive_Beamforming}.
\begin{tabular}{@{} >{\centering}b{251pt}}
\setlength{\parindent}{10pt}
Let$\hspace{0.06cm}$ $h_i$ $\hspace{0.05cm}$denotes $\hspace{0.04cm}$the $\hspace{0.05cm}$small-scale$\hspace{0.07cm}$ fading channel $\hspace{0.03cm}$coefficient 
\end{tabular}
 between the destination (D) and \textit{$i^{th}$} element of the RIS, we have $h_i\sim \mathcal{CN}(0,1)$ under Rayleigh fading assumption, where $\mathcal{C}\mathcal{N}(0,\sigma^2)$ stands for complex Gaussian distribution with zero mean and $\sigma^2$ variance. $\theta_0$ and $\theta_1$ stand for the phases of two \textit{M}-PSK symbols to be transmitted according to $2\log_{2}(M)$ bits. Assuming quasi-static fading channels, where the channels will remain constant over the two time slots, the received signal at the first time slot can be written as\\
 \vspace{-0.3cm}
\setlength{\abovedisplayskip}{3pt}
\setlength{\belowdisplayskip}{3pt}
\begin{align}\label{r_0 detailed form}
r_0&= \sqrt{P_L}\left[\sqrt{E_s}e^{j\theta_0}\sum_{i=1}^{N/2}{h_{i}}+\sqrt{E_s}e^{j\theta_1}\hspace{-2mm} \sum_{i=\frac{N}{2}+1}^{N}\hspace{-0.1cm}{h_{i}}\right]+n_0
\end{align}\\
where $n_0$ is the additive white Gaussian noise (AWGN) sample at the first time slot, $n_0\sim\mathcal{C}\mathcal{N}(0,\textit{N}_0)$. $E_s$ is the transmitted RF signal energy and $\theta_0$ and $\theta_1$ are the common RIS reflection phases for the first and second parts, respectively, for the first time slot. $P_L$ is the total path gain (loss), and more details regarding the considered path loss model and environmental setups will be given in Section IV. According to the Alamouti's transmission scheme, in the second time slot, we obtain the following received signal by carefully adjusting the common RIS phase terms as $-(\theta_1+\pi)$ and $-\theta_0$ for the first and second parts, respectively:\\
\vspace{-0.3cm}
\setlength{\abovedisplayskip}{3pt}
\setlength{\belowdisplayskip}{3pt}
\begin{align}
r_1&= \sqrt{P_L}\left[\sqrt{E_s}e^{-\textit{j}(\theta_1+\pi)} \sum_{\textit{i}=1}^{\textit{N/2}}{h_{\textit{i}}}+\sqrt{E_s}e^{-\textit{j}\theta_0} \hspace{-2mm}\sum_{\textit{j}=\frac{N}{2}+1}^{\textit{N}}{h_{\textit{i}}}\right]+n_1 
\end{align}
where $n_1\sim\mathcal{C}\mathcal{N}(0,\textit{N}_0)$. Defining $s_{0}=\sqrt{E_s}e^{\theta_0}$, $s_{1}=\sqrt{E_s}e^{\theta_1}$, $A_{0}=\sqrt{P_L}\sum_{\textit{i}=1}^{\textit{N/2}}{h_{\textit{i}}}$ and $A_{1}=\sqrt{P_L}\sum_{\textit{i}=\frac{N}{2}+1}^{\textit{N}}{h_{\textit{i}}}$, (1) and (2) can be re-expressed as
\begin{align}\label{r_0 and r_1 compact form}
& r_0=s_0 A_0+s_1 A_1+n_0\\
& r_1=-s_1^{*} A_0+s_0^{*} A_1+n_1
\end{align}
where ${s_0}$ and ${s_1}$ stand for two virtual $M$-PSK symbols to be delivered to the receiver and $M$ is the modulation order. As in the classical Alamouti's scheme, the combiner will construct the combined signals as follows: 
\begin{align}
\tilde{s_0}  =r_0 A_0^{*}+r_1^{*} A_1=(|A_0|^2+|A_1|^2)s_0+A_0^{*} n_0+A_1 n_1^*,
\end{align}
\begin{align}
\tilde{s_1} =r_0 A_1^{*}-r_1^{*} A_0=(|A_0|^2+|A_1|^2)s_1-A_0 n_1^{*}+A_1^{*} n_0.
\end{align}
Then, $\tilde{s}_0$ and  $\tilde{s}_1$ will be passed to the maximum likelihood (ML) detector to estimate $s_0$ and $s_1$.
Considering the symmetry of $s_0$ and $s_1$, the instantaneous received SNR per symbol can be obtained as
\begin{align}\label{SNR}
\gamma=\frac{(|A_0|^2+|A_1|^2)E_s}{\textit{N}_0}.
\end{align}
Considering $h_i\sim \mathcal{CN}(0,1)$, we obtain $A_0$ and $A_1\sim \mathcal{CN}(0,P_L\frac{\textit{N}}{2})$. Consequently, $\gamma$ becomes a central chi-square distributed random variable with four degrees of freedom with the following MGF \cite{Proakis}:
\begin{align}\label{MGF of SNR}
M_{\gamma}(s)=\left(\frac{1}{1-{\frac{sP_LNE_s}{2N_0}}}\right)^2.
\end{align}
 From (8), the average symbol error probability (SEP) for \textit{M}-PSK signaling can be obtained as \cite{Alaouni}
\begin{align}\label{SEP for M-PSK}
\hspace{-1cm}P_e=\frac{1}{\pi}\int_{0}^{(M-1)\pi/M}\;\left(\frac{1}{1+\left(\frac{\sin{(\pi/M)^2}}{\sin{(\eta)}^2}\right)\frac{P_LNE_s}{2N_0}}\right)^2\textit{d}\eta
\end{align}
which can be simplified for BPSK as
\begin{align}\label{BER for BPSK}
P_e=\frac{1}{\pi}\int_{0}^{\pi/\textit{2}}\;\left(\frac{1}{1+\frac{P_LNE_s}{2N_0\sin{(\eta)}^2}}\right)^2\textit{d}\eta.
\end{align}
From (9), we observe that a transmit diversity of order two is still achieved by this scheme due to the fact that the RIS mimics a similar  transmission methodology to the Alamouti's scheme over two time slots and preserves its orthogonality. Furthermore, the transmitted symbols have an SNR amplification achieved by the combination of the signals reflected from the RIS, where the SNR is enhanced by a factor of \textit{N}, as seen from (10).
\section{RIS-Assisted and IM-Based VBLAST SYSTEM: The signal model}\label{sec3}
The proposed RIS-assisted and IM-based VBLAST scheme is assumed to
utilize the RIS through a feedback link to eliminate the channel phases between a transmit-receive antenna pair. Instead of randomly selecting it, this transmit-receive antenna pair is selected according to the bits incoming to the RIS from S through the feedback link, in IM fashion. This means that the proposed scheme benefits from the RIS in an effective way. Consequently, with the help of the RIS, the proposed scheme provides a significant BER performance enhancement and an effective boost in the spectral efficiency compared to the classical VBLAST scheme. The proposed scheme is shown in Fig. 2, where an $N_t\times N_r$ VBLAST system is being 
 \begin{figure}[t]
	\begin{center}
		\includegraphics[width=9cm]{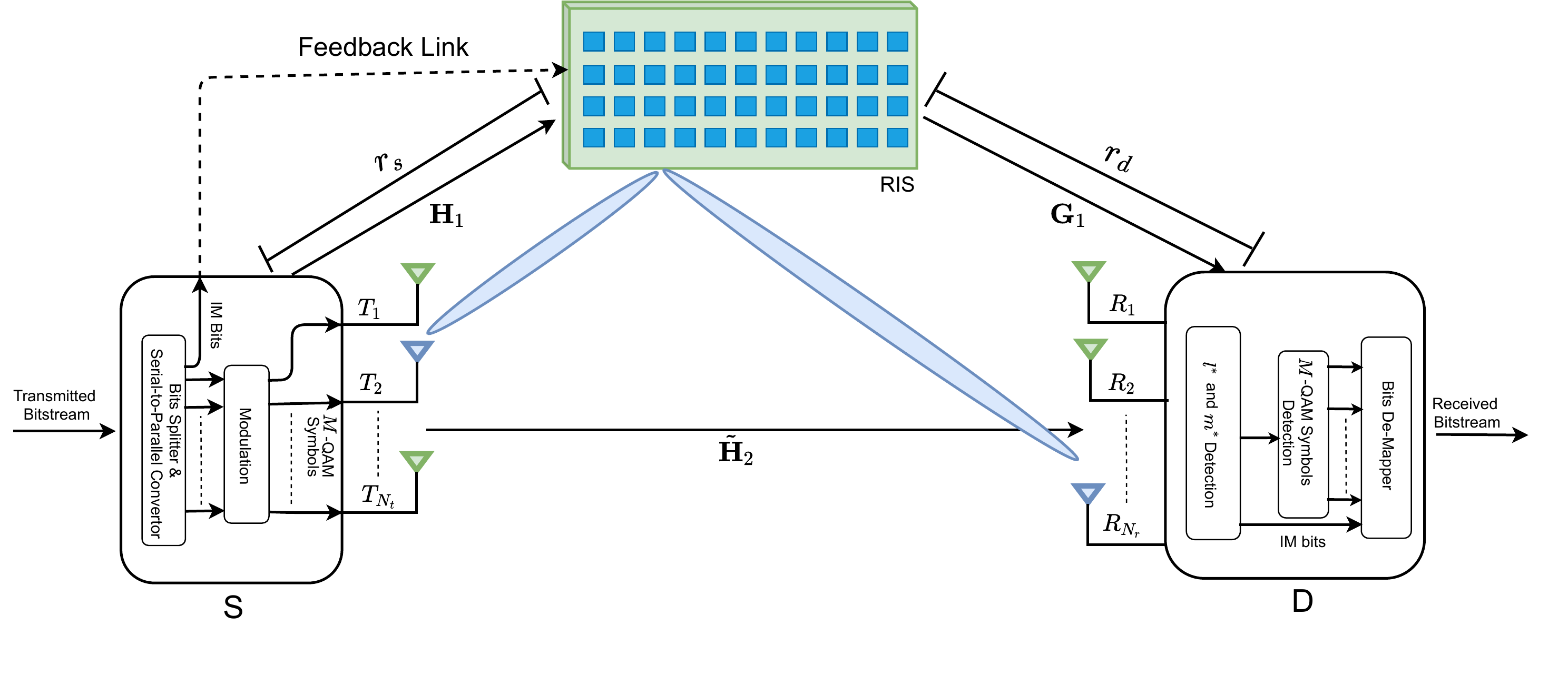}
		\caption{RIS-assisted and IM-based $N_t\times N_r$ VBLAST scheme.}
	\end{center}
	\vspace{-0.4cm}
\end{figure}
\noindent operated along with an RIS with $N$ reflecting elements. The RIS-assisted and IM-based VBLAST scheme can be operated in three different modes namely, i) full IM mode, ii) partial-IM mode, and iii) enhancing mode. In what follows, we describe these three operating modes.

 In the full-IM mode, at the transmitter side, the IM-based mapping and transmission can be described as follows. The incoming bits will be divided into two groups, the first group of $N_t\log_{2}M$ bits will be used to select $N_t$ independent $M$-QAM symbols to be transmitted from the available $N_t$ transmitting antennas at S, as in classical VBLAST scheme. 

 The second group of $\log_{2}\left(N_t N_r\right)$ bits, where $N_tN_r$ is assumed to be an integer power of two and corresponds to the all possible transmit-receive antenna combinations, is sent through a feedback link from S to the RIS. The RIS uses these incoming bits to select the indices of the transmit and receive antennas, shown by $l^*$ and $m^*$, corresponding to a $T_{l^*}-R_{m^*}$ pair of antennas, respectively. According to the CSI associated with the selected antenna pair, the phase shifts of the RIS elements will be adjusted to make the $T_{l^*}$-RIS-$R_{m^*}$ equivalent channel phases equal to zero. Consequently, the signals transmitted from $T_{l^*}$ and reflected from the RIS will constructively combine to provide SNR amplification at $R_{m^*}$. The overall spectral efficiency of the system becomes $N_t\log_{2}M+\log_{2}\left(N_t N_r\right)$ bits per channel use (bpcu).
 
 In the partial-IM mode, the transmission procedure is the same as in the full-IM mode except that a smaller set of the transmit-receive antenna combinations is used to convey the IM bits. That is, the index of the targeted transmit antenna is determined first and the same index is used for the targeted receive antenna, $T_{l^*}-R_{l^*}$. Hence, there are only $\log_{2}N_t$ possible combinations that can be used to convey the IM bits. The resulting spectral efficiency under this mode will be $N_t\log_{2}M+\log_{2}N_t$ bpcu, where the motivation here is to sacrifice spectral efficiency gained by IM in order to further enhance the BER performance through increasing the reliability of IM bits.
  
   Finally, in the enhancing mode, IM is not performed at all and therefore, there is no need for a feedback link between S and D, instead, a fixed and predefined antenna pair will always be  targeted by the RIS for the elimination of channel phases. Consequently, the best BER performance is achieved while preserving the same spectral efficiency for classical VBLAST, $N_t\log_{2}M$. For all operating modes, we assume that perfect CSI is available at the both RIS and receiver side. The acquisition of CSI in RIS-based systems is discussed in \cite{RIS-ESTIMATION}, \cite{CSI_RIS1}, \cite{CSI_RIS3}. We describe our scheme with an example.

\textit{Example}: A $4\times4$ RIS-assisted VBLAST system operated in the full-IM mode with QPSK modulation transmits the bitstream of $[00\;01\;10\;11\;00 \;01]$ as follows. The first $8$ bits are modulated to $4$ QPSK symbols and transmitted, in parallel, from the available $4$ transmit antennas. The remaining $4$ bits are used by the RIS to select the pair $T_1-R_2$, since we implement the following mapping rule:  $00\rightarrow1$ and $01\rightarrow2$. The RIS adjusts the reflection phases to eliminate the channel phases between $T_1$ and $R_2$. Here, we assume that the adjustment of the phases of the RIS elements and the $M$-QAM symbols' transmission from S are being performed simultaneously. On the other hand, the receiver tries  to first estimate the indices of the selected antennas, and then successively detects the $N_t$ independent $M$-QAM symbols.

 For an $N_t\times N_r$ VBLAST system assisted by an RIS with $N$ reflectors, the vector of the received signals 
$\mathbf{r}\in\mathbb{C}^{N_r\times1}$ can be written as \cite{Enrg_efficiency}
\setlength{\abovedisplayskip}{5pt}
\setlength{\belowdisplayskip}{5pt}
\begin{align}\label{recieved vector VBLAST}
\mathbf{r}=\left[\sqrt{P_{L1}}\mathbf{ {G}}_1^T\mathbf{\Theta}\mathbf{ H}_1 +\sqrt{P_{L2}}\tilde{\mathbf{H}}_2\right]\mathbf{x}+\mathbf{n}=\mathbf{V}\mathbf{x}+\mathbf{n}
\end{align}
where $\mathbf{H}_1\in\mathbb{C}^{N\times N_t}$, $\mathbf{G}_1\in\mathbb{C}^{N\times N_r}$ are the S-RIS and RIS-D, uncorrelated Rician fading channel matrices, respectively. The S-D channel matrix $\tilde{\mathbf{H}}_2\in\mathbb{C}^{N_r\times N_t}$ is a random matrix where its elements are independent and identically distributed (i.i.d) complex Gaussian random variables (RVs) with zero mean and unit variance, $\sim\mathcal{C}\mathcal{N}(0,1)$. $K$ is the Rician factor standing for the ratio of LOS and NLOS power and $\mathbf{\Theta}=\mathrm{diag}(e^{j\Phi_1}, ... ,  e^{j\Phi_n}, ... , e^{j\Phi_N})$ is the matrix of RIS reflection phases. In theory, each RIS element can be adjusted to any arbitrary phase shift, $\Phi_n\in[0,2\pi)$, however, this is challenging to implement in practice. Therefore, in implementation, the RIS elements are being designed so that they can be adjusted to a finite number of discrete phase shifts. Hence, assuming uniform quantization for the interval $ [0,2\pi)$, the phase shift associated with each RIS element can be controlled by $b$ bits that correspond to $Z=2^b$ possible different phase shifts belong to the finite set $\mathcal{F}=\{0, \Delta \Phi, ..., \Delta \Phi(Z-1)\}$, where $\Delta \Phi=\frac{2\pi}{Z}$ \cite{BeamformingPhs}. $\mathbf{x}\in\mathbb{C}^{N_t\times 1}$ is the vector of transmitted $M$-QAM symbols, $\mathbf{V}\in\mathbb{C}^{N_r\times N_t}$ is the S-RIS-D and S-D equivalent channel matrix and  $\mathbf{n}\in\mathbb{C}^{N_r\times 1}$ is the vector of AWGN noise samples. $P_{L1}$ and $P_{L2}$ are the total path losses for S-RIS-D and S-D transmission paths, respectively, and more details are given in Section IV.
The received signal by the $m^{th}$ receiving antenna ($R_m$) can be represented as
\begin{align}\label{recieved signal by antenna VBLAST}
r_m=\left[\sqrt{P_{L1}}\mathbf {g}_{m}^{T}\mathbf{\Theta} \mathbf{H}\mathbf+\mathbf{h}_2\right]\mathbf{x} +n_m
\end{align}
where $\mathbf{h}_2=\sqrt{P_{L2}}\tilde{\mathbf{h}}_2$ and $\tilde{\mathbf{h}}_2\in\mathbb{C}^{1\times N_t}$ is the S-D channel vector for the $m^{th}$ receiving antenna. $\mathbf{g}_m\in\mathbb{C}^{N\times 1}$ is the RIS-D channel vector for the $m^{th}$ receiving antenna, and $n_m$ is the AWGN sample with $n_m\sim\mathcal{C}\mathcal{N}(0,\textit{N}_0)$.
Expanding (12), we have
\setlength{\abovedisplayskip}{6pt}
\setlength{\belowdisplayskip}{6pt}
\begin{align}
r_m=\sqrt{P_{L1}}\sum_{l=1}^{N_t}\left[\sum_{i=1}^{N}h_{i}^{(l)}e^{j\Phi_i}g_{i}^{(m)}\right]x^{(l)}+\mathbf{h}_2\mathbf{x}+n_m
\end{align}\\
\begin{tabular}{@{} >{\centering}b{252pt}}
	 where $h_{i}^{(l)}\hspace{-0.05cm}=\alpha_{i}^{l}e^{-j\theta_{i}^{(l)}}\hspace{-0.05cm}$ is the S ($l^{th}$ transmitting antenna)-RIS\\
	\vspace{-0.1cm}
	($i^{th}$element) channel coefficient and $g_{i}^{(m)}\hspace{-0.1cm}=\hspace{-0.05cm}\beta_{i}^{m}e^{-j\Psi_{i}^{(m)}}\hspace{-0.1cm}$ is RIS\\
	($i^{th}$element)-D $\hspace{-0.05cm}$($m^{th}\hspace{-0.08cm}$ receiving antenna) channel coefficient. In
\end{tabular} this way, in order to eliminate the phases of the S-RIS-D channel between the antenna pair $T_{l^*}-R_{m^*}$,  elements' phases of the RIS are adjusted as $\Phi_i=\theta_{i}^{(l^*)}+\Psi_{i}^{(m^*)}$, and (13) can be re-expressed for the $(m^*)^{th}$ receive antenna as
\begingroup
\small
\begin{flalign}\label{recieved signal by antenna VBLAST}
r_{m^{*}}\hspace{-0.5mm}&=\hspace{-1mm}\sqrt{P_{L1}}\left[\left[\sum_{i=1}^{N}\alpha_{i}^{l^*}\beta_{i}^{m^*}\right]{x}^{(l^*)}+\hspace{-0.4cm}\sum_{l=1, l\neq l^*}^{N_t-1}\hspace{-0.5mm}\left[\sum_{i=1}^{N}h_{i}^{(l)}e^{j\Phi_i}g_{i}^{(m^*)}\right]{x}^{(l)}\right]\nonumber\\ &+\mathbf{h}_2\mathbf{x}+n_{m^*}.
\end{flalign}
\endgroup
(14) can be interpreted as follows. The first term illustrates the amplification gained by the constructive combining of the  signals reflected from the RIS and belongs to the symbol $x^{(l^*)}$. This constructive combining will result in an SNR gain of $N^2$ for this symbol as in \cite{Transmission_conference}. Hence, the BER performance of this symbol will be boosted up and consequently, it will be the strongest symbol where the nulling and canceling algorithm starts with. This also means that the error propagation from the first symbol to the remaining ones will be significantly mitigated and the overall BER performance will be improved. The second term shows the destructive interference of the signals reflected from the RIS, which belongs to the other symbols. Finally, the third term corresponds to the interference received by the $(m^*)^{th}$ receiving antenna through the S-D transmission path for all the transmitted symbols. Hence, compared to the classical VBLAST, an RIS will introduce $N$ times extra interference for each received symbol. Nevertheless, assuming that the CSI over S-RIS-D is available at the receiver side, then this interference can be handled readily.
\subsection{Detection Algorithms}
At the receiver side, we introduce two novel nulling-based detectors to detect the transmit-receive antenna indices targeted by the RIS for channel phases elimination. According to Algorithm 1, the optimal detector performs an exhaustive search for $l^*$ and $m^*$ jointly, as follows. For each iteration, the detector determines $\mathbf{\Theta^{(\textit{l},\textit{m})}}$ and then constructs $\hat{\mathbf{V}}^{(\textit{l},\textit{m})}$. Next, the nulling-based procedure will be used to detect the first symbol, which has the highest SNR, assumed to be transmitted and received by the antenna pair $T_l-R_m$. Finally, the pair $l$ and $m$  associated with the symbol that has the minimum squared Euclidean distance will be picked as the pair $l^*$ and $m^*$.

In Algorithm 1, $\hat{\mathbf{V}}^{(l,m)}$ is the index-estimated S-RIS-D and S-D equivalent channel matrix assuming the antenna pair $T_{l}-R_{m}$ was targeted by the RIS for channel phases elimination. $\mathbf{\Theta}^{(l,m)}$ is the diagonal RIS phases matrix where its $i^{th}$ element $\Phi_i=\theta_i^{(l)}+\Psi_i^{(m)}$ corresponds to the phase elimination for the S-RIS-D channel between $T_{l}$ and $R_{m}$. ${(\cdot)^+}$ is Moore-Penrose pseudo-inverse operator, $(\hat{\mathbf{V}}^{(l,m)})^+=(\hat{\mathbf{V}}^{(l,m)})^{\dag} (\hat{\mathbf{V}}^{(l,m)} ({\hat{\mathbf{V}}^{(l,m)}})^\dag)^{-1}$, where ${(\cdot)}^{\dag}$ is the Hermitian operator. $(\mathbf{W}^{(l,m)})_j$ is the $j^{th}$ row of $\mathbf{W}^{(l,m)}$. $k_l$ is the index of the row, which has the minimum squared Euclidean norm, of the matrix $\mathbf{W}^{(l,m)}$, corresponding to the symbol with the highest SNR. $(\textbf{W}^{(l,m)})_{k_l}$ is the $k_{l}^{th}$ row of $\mathbf{W}^{(l,m)}$, $y_{k_l}$ is the $k_{l}^{th}$ symbol after nulling the interference of the other symbols. $\mathcal{Q}(\cdot)$ returns the squared Euclidean  distance, $D_{l,m}$, of the closest $M$-QAM symbol associated with $y_{k_l}$. Hence, $\hat{\mathbf{V}}^{(\hat{l},\hat{m})}$ associated with the minimum distance $D_{\hat{l},\hat{m}}$ is the most likely the equivalent channel matrix that corresponds to the current adjustment of the RIS phases.  
\begin{algorithm}[h]
	\label{detctorA}
	\caption{Optimal detector: Detecting the antenna indices $l^*$ and $m^*$ jointly.}
	\begin{algorithmic}[1]
		\REQUIRE $\mathbf{H}_1$,  $\tilde{\mathbf{H}}_2$ $\mathbf{G}_1$, $\mathbf{r}$, $\sqrt{P_{L1}}$, $\sqrt{P_{L2}}$
		\FOR{$l=1:N_t$}
		\FOR{$m=1:N_r$}
		\STATE $\hat{\mathbf{V}}^{(l,m)}=\sqrt{P_{L1}}\mathbf{ {G}_1}^T\mathbf{\Theta}^{(l,m)}\mathbf{H}_1+\sqrt{P_{L2}}\tilde{\mathbf{H}}_2$
		\STATE$\mathbf{W}^{(l,m)}=(\hat{\mathbf{V}}^{(l,m)})^+$ 
		\STATE $k_l= \underset{j\in\{1, 2, ..N_t\}}{\textrm{arg min}}\;\ \big | \big |(\mathbf{W}^{(l,m)})_j{\big | \big |}^2$
		\COMMENT{Ordering}
		\STATE $\textbf{s}_{k_l}=(\textbf{W}^{(l,m)})_{k_l}$
		\STATE $y_{k_l}=\textbf{s}_{k_l}^T \mathbf{r}$
		\hspace{3.2cm}\COMMENT{Nulling}
		\STATE $D_{l,m}=\mathcal{Q}(y_{k_l})$
		\ENDFOR
		\ENDFOR 
		\STATE	$\{\hat{l},\hat{m}\}=\underset{{\substack{l\in\{1, 2, ..N_t\}\\ m\in\{1, 2, ..N_r\}}}}{\textrm{arg\;min}}\;D_{l,m}$
		\STATE \hspace{0.0cm} $\hat{\mathbf{V}}=\sqrt{P_{L1}}\mathbf{ {G}_1}^T\mathbf{\Theta}^{(\hat{l},\hat{m})}\mathbf{H}_1+\sqrt{P_{L2}}\tilde{\mathbf{H}}_2$
		\vspace{0.1cm}
		\RETURN $\hat{\mathbf{V}}$
	\end{algorithmic}
\end{algorithm}
By estimating $l^*$ and $m^*$, the IM bits conveyed by the  adjustment of the RIS phases  will be obtained. In order to reduce the complexity of the detector proposed in Algorithm 1, the receiving antenna index, $m^*$, can be detected using a greedy detector instead of the joint exhaustive search for $l^*$ and $m^*$. In this way, $m^*$ can be detected by finding the receiving antenna with the highest instantaneous energy:
\begin{flalign}\label{Greedy}
\hspace{0.5cm}\hat{m}=\underset{m\in\{1, 2, . . .N_r\}}{\textrm{arg max}}|r_m|^2
\end{flalign}
Next, a nulling-based procedure is used to search for $l^*$ while fixing $\hat{m}$ found from (15). This detector is represented as a sub-optimal one and Algorithm 2 illustrates its detection steps.
\setlength{\abovedisplayskip}{1pt}
\setlength{\belowdisplayskip}{3pt}

After obtaining the S-RIS-D and S-D equivalent channel matrix, $\hat{\mathbf{V}}$, it will be used by Algorithm 3 to detect the $N_t$ independent symbols as in the case of classical VBLAST \cite{VBLAST}. In Algorithm 3, $\tilde{\mathcal{Q}}(\cdot)$ is the slicing function, $(\hat{\mathbf{V}})_{k_i}$ is the $k^{th}_i$ column of $\hat{\mathbf{V}}$ and $({\hat{\mathbf{V}}_{\overline{\textit{k}_{\textit{i}}}}})^+$ is pseudo inverse of the matrix obtained by zeroing the columns of $\hat{\mathbf{V}}$ with indices $k_1$, $k_2$, ...$k_i$. Finally, $\mathbf{r}_{i+1}$ is the signal vector after subtracting the interference contribution of the previously detected symbol $\hat{x}_{k_i}$. 
\vspace{0.3cm}
\subsection{Computational Complexity Analysis}
At the receiver side, first, the RIS-assisted VBLAST scheme uses Algorithm 1 or Algorithm 2 to detect the indices $l^*$ and $m^*$. Second, the receiver uses Algorithm 3 to detect the $N_t$ transmitted $M$-QAM symbols. Here, we calculate the computational complexity associated with Algorithms 1, 2, and 3.\\
\vspace{-0.05cm}
Denoting the total number of complex multiplications (CMs) required by Algorithms 1, 2, and 3 as $C_1$, $C_2$, and $C_3$, respectively, then from Table 1, $C_1$, $C_2$ and $ C_3$ can be calculated as follows
\begin{align}
\hspace{0cm}C_1 &=N_tN_r[(N_rN+N_rNN_t)+(2N_r^2N_t+N_r^3)+N_tN_r\nonumber\\&+N_r+2^M],\\
\hspace{0cm}C_2 &=N_r+N_t[(N_rN+N_rNN_t)+(2N_r^2N_t+N_r^3)+N_tN_r\nonumber\\ &+N_r+2^M],\\ \nonumber
\end{align}
\vspace{-0.5cm}
\begin{algorithm}[h]
	\label{detctorA}
	\caption{Sub-optimal detector: Detecting the antenna indices $l^*$ and $m^*$ sequentially.}
	\begin{algorithmic}[1]
		\REQUIRE $\mathbf{H}_1$,  $\tilde{\mathbf{H}}_2$ $\mathbf{G}_1$, $\mathbf{r}$, $\sqrt{P_{L}}$, $\sqrt{P_{L2}}$
		\STATE $\hat{m}=\underset{{\substack{m\in\{1, 2, ..N_r\}}}}{\textrm{arg\;max}}\;\;|r_m|^2$
		
		%
		%
		\FOR{$l=1:N_t$}
		\STATE $\hat{\mathbf{V}}^{(l,\hat{m})}=\sqrt{P_{L1}}\mathbf{ {G}_1}^T\mathbf{\Theta}^{(l,\hat{m})}\mathbf{H}_1+\sqrt{P_{L2}}\tilde{\mathbf{H}}_2$
		\STATE$\mathbf{W}^{(l,\hat{m})}=(\hat{\mathbf{V}}^{(l,\hat{m})})^+$ 
		\STATE $k_l= \underset{j\in\{1, 2, ..N_t\}}{\textrm{arg min}}\;\ \big | \big |(\mathbf{W}^{(l,\hat{m})})_j{\big | \big |}^2$
		\COMMENT{Ordering}
		\STATE $\textbf{s}_{k_l}=(\textbf{W}^{(l,\hat{m})})_{k_l}$
		\STATE $y_{k_l}=\textbf{s}_{k_l}^T \mathbf{r}$
		\hspace{3.2cm}\COMMENT{Nulling}
		\STATE $D_l=\mathcal{Q}(y_{k_l})$
		\ENDFOR
		\STATE $\hat{l}=\underset{{\substack{l\in\{1, 2, ..N_t\}}}}{\textrm{arg\;min}}\;D_{l}$
		\STATE \hspace{-0.15cm} $\hat{\mathbf{V}}=\sqrt{P_{L1}}\mathbf{ {G}_1}^T\mathbf{\Theta}^{(\hat{l},\hat{m})}\mathbf{H}_1+\sqrt{P_{L2}}\tilde{\mathbf{H}}_2$
		\vspace{0.1cm}
		\RETURN $\hat{\mathbf{V}}$
	\end{algorithmic}
\end{algorithm}
\vspace{-0.2cm}
\begin{algorithm}[h]
	\label{detctorA}
	\caption{ZF-based successive nulling and canceling to detect the transmitted $M$-QAM symbols.}
	\begin{algorithmic}
		\REQUIRE $\mathbf{\hat{V}}$, $\mathbf{r}$
		\STATE $\mathbf{r}_1=\mathbf{r}$
		\STATE $\mathbf{W_1}=\mathbf{\hat{V}}^+$
		\STATE $k_1= \underset{j\in\{1, 2, ..N_t\}}{\textrm{arg min}}\;\ \big | \big |(\mathbf{W}_1)_j{\big | \big |}^2$
		\FOR{$i=1:N_t-1$}
		\STATE $\textbf{s}_{k_i}=(\textbf{W}_i)_{k_i}$
		\STATE $y_{k_i}=\textbf{s}_{k_i}^T \mathbf{r}_i$
		\STATE $\hat{{x}}_{k_i}=\tilde{\mathcal{Q}}(y_{k_i})$
		\hspace{2cm}\COMMENT{Slicing}
		\STATE $\mathbf{r}_{i+1}=\mathbf{r}_i-\hat{x}_{k_i}(\hat{\mathbf{V}})_{k_i}$
		\hspace{0.8cm}\COMMENT{Canceling}
		\STATE $\mathbf{W}_{i+1}=({\hat{\mathbf{V}}_{\overline{\textit{k}_{\textit{i}}}}})^+$ 
		\STATE $k_{i+1}=\underset{j\in\{1,..., N_t\}\setminus
			\{k_1, . . ., k_{i}\}}{\textrm{arg min}}\big | \big |(\mathbf{W}_{i+1})_j{\big | \big |}^2$
		\ENDFOR
		\STATE $\textbf{s}_{k_{i+1}}=(\textbf{W}_{i+1})_{k_{i+1}}$
		\STATE $y_{k_{i+1}}=\textbf{s}^T_{k_{i+1}} \mathbf{r}_{i+1}$
		\STATE $\hat{x}_{k_{i+1}}=\tilde{\mathcal{Q}}(y_{k_{i+1}})$
		
		\RETURN $\mathbf{\hat{x}}$
	\end{algorithmic}
\end{algorithm}
\setlength{\abovedisplayskip}{3pt}
\setlength{\belowdisplayskip}{3pt}
\begin{align}
C_3 &=(2N_r^2N_t+N_r^3)+N_tN_r+(N_t-1)[N_r+2^M+N_r\nonumber\\ 
&+(2N_r^2N_t+N_r^3)+N_tN_r]+N_r+2^M.
\end{align}
In order to provide a useful insight on the overall complexity, let $N_t=N_r$. Then (16), (17), and (18)  can be rewritten as
\begin{align}
C_1=&N(N_r^3+N_r^4)+3N_r^5+N_r^4+N_r^3+N_r^22^M,\\
\hspace{-0.3cm}C_2 =&N_r+N(N_r^2+N_r^3)+3N_r^4+N_r^3+N_r^2+N_r2^M,\\
C_3 =&4N_r^3+3N_r^2+N_r2^M+3N_r^4+N_r+2^M.
\end{align} 
From (16) and (17), it can be seen that the number of antennas and the number of RIS reflecting elements both have a significant contribution to the computational complexity of detecting the antenna indices $l^*$ and $m^*$. This is still valid even if we consider the fact that $N$ is, practically, much larger than $N_r$, since the latter still has a higher exponent. From (19) and (20), the overall computational complexity level can be obtained as $\sim \mathcal{O}(N(N_r^3+N_r^4)+N_r^5)$, and $\sim \mathcal{O}(N(N_r^2+N_r^3)+N_r^4)$, for Algorithms 1 and 2, respectively. Comparing both algorithms, we observe that the overall computational complexity is dominated by that of constructing $\hat{\mathbf{V}}^{(l,m)}$. Compared to the classical VBLAST scheme, where the receiver computational complexity will be equivalent to that of Algorithm 3 only, $\sim \mathcal{O}(N_r^4)$, RIS-assisted IM-based VBLAST scheme has an extra complexity of $\sim \mathcal{O}(N(N_r^3+N_r^4))$ and $\sim \mathcal{O}(N(N_r^2+N_r^3))$ for Algorithms 1 and 2, respectively.
\begin{table}[h]
	\caption{Computational complexity derivation steps of nulling-based  detection algorithms.}
	\label{tab:table-name}
	\begin{center}		
		\begin{tabular}{ | m{8em} | m{12em}| } 
			\hline
			\textit{Operation} & \textit{CMs} \\ 
			\hline
			Searching for $\hat{m}$ & $N_r$ \\ 
			\hline
			Constructing $\hat{\mathbf{V}}^{(l,m)}$  & $N_rN$+$N_rNN_t$ \\ 
			\hline
			Pseudo inverse of $\hat{\mathbf{V}}^{(l,m)}$ & $2N_r^2N_t+N_r^3$\\
			\hline
			Ordering & $N_tN_r$\\
			\hline
			Nulling & $N_r$\\
			\hline
			Getting $D_{l,m}$ & $2^M$\\
			\hline
		\end{tabular}
	\end{center}
	\vspace{-0.4cm}
\end{table}
 The cost of this additional complexity stems for the construction of $\hat{\mathbf{V}}^{(\hat{l},\hat{m})}$, which is required to detect the transmitted symbols and the indices of the targeted transmit-receiving antennas.
In a brief, $C_1$ or $C_2$ is the additional computational complexity that is required to operate a classical VBLAST system as an RIS-assisted  system.


\section{Simulation results}
In this section, exhaustive computer simulations are provided for the proposed schemes against their counterparts. We consider realistic setups and path loss models where both transmitter and receiver located in an indoor environment and Fig. 3 shows the block diagrams of the benchmark schemes. In all simulations, the SNR is defined to be $E_s/N_0$. Furthermore, perfect CSI is assumed to be available for the proposed and benchmark schemes at the receiver side only, except RIS-assisted and IM-based VBLAST scheme where the CSI need also to be available at the RIS side.

Figs. 3 (a) and (c) show the system setups for the classical Alamouti and VBLAST schemes, where the S-D channel is assumed to follow a Rayleigh fading (i.e., consist of a non-LOS link). For the RIS-AP scheme shown in Fig. 3 (b), the S-D transmission path is assumed to be fully blocked by obstacles and the only transmission path is through the RIS, where the S-RIS path is LOS dominated and the RIS-D path follow Rayleigh fading \cite{Transmission_conference}.

 For classical VBLAST and Alamouti schemes, where there is no RIS,  the path loss is calculated for an operating frequency 
of $1.8$ GHz, as follows \cite{Cost231}
\begin{align}
&P_L^{(S-D)}(R)[\mathrm{dB}]=42.7+20\log_{10}R+13.8 \label{16}
\end{align}
where $R$ is the S-D separation distance, $42.7$ dB is the path loss at one meter distance, and $13.8$ dB corresponds to the path loss of two walls each having $6.9$ dB of path loss.

For the RIS-assisted Alamouti and RIS-AP schemes, the S-RIS-D path loss, $P_L$, is calculated as follows \cite{Ellingson}
\setlength{\abovedisplayskip}{3pt}
\setlength{\belowdisplayskip}{5pt}
\begin{align}
P_L^{(S-RIS-D)}=\frac{\lambda^4}{256\pi^2r_s^2r_d^2}
\end{align}
where $\lambda$ is the wavelength of the operating frequency ($1.8$ GHz). Finally, for RIS-assisted and IM-based VBLAST scheme, $P_{L2}$ and $P_{L1}$ are calculated from (22) and (23), respectively. For RIS-assisted Alamouti, RIS-AP, and classical 
\begin{figure}[t] 
	\begin{center}
		\includegraphics[width=3.5in, height=2.7in]{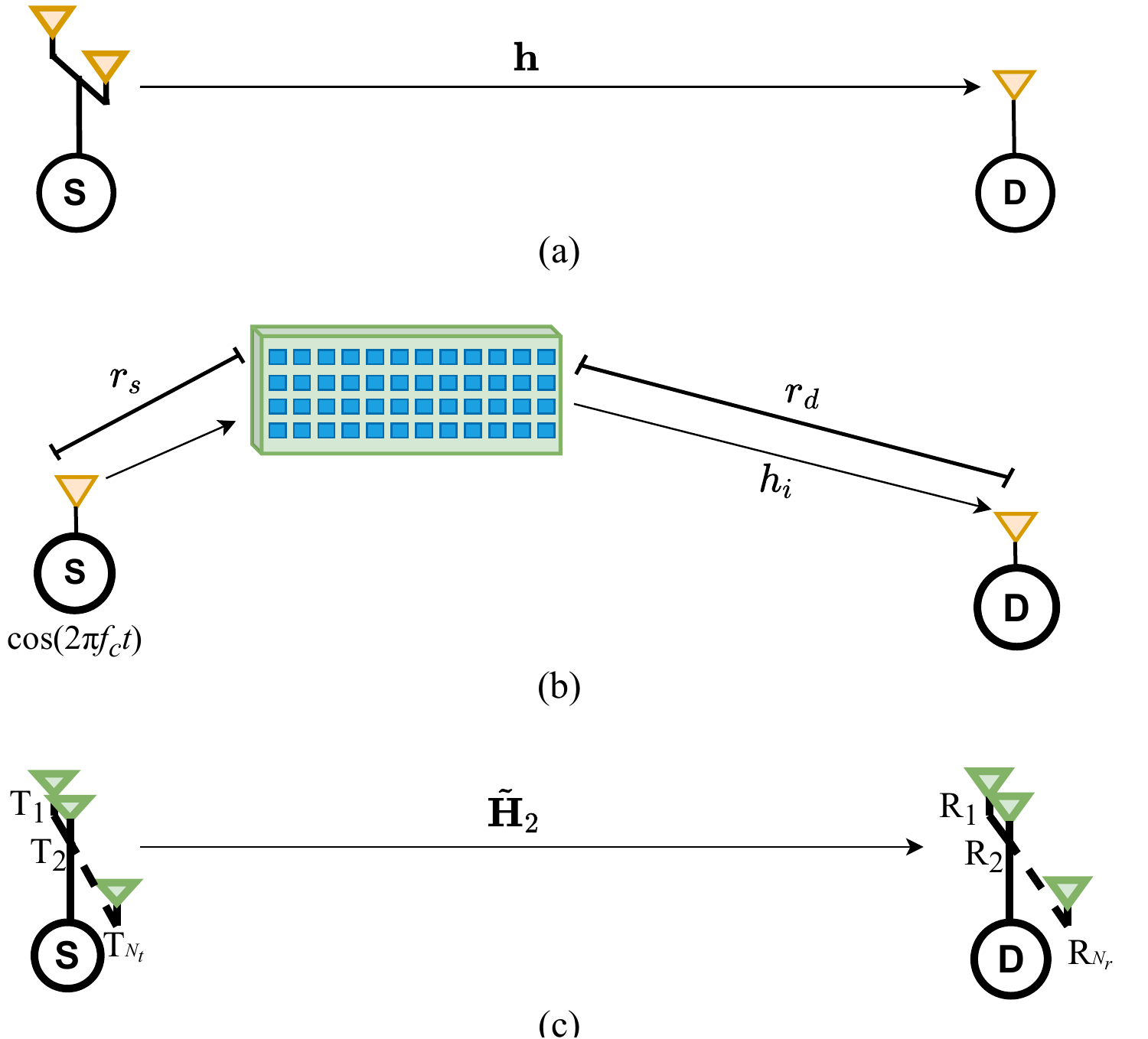}
		\caption{(a) Classical Alamouti's scheme. (b) RIS-AP scheme.
			(c) Classical VBLAST scheme.}
	\end{center}
	\vspace{-0.5cm}	
\end{figure}
Alamouti's schemes we have $r_s=1\;\text{m}$, $r_d=9\;\text{m}$, $b=0.5\;\text{m}$, and $R=9.85\;\text{m}$. On the other hand, for RIS-assisted and IM-based VBLAST and classical VBLAST schemes, we
have $r_s=3\;\text{m}$, $r_d=3\;\text{m}$, $b=0.5\;\text{m}$, and $R=5.91\;\text{m}$.

  Fig. 4 illustrates the BER performance of the RIS-assisted Alamouti's scheme versus the classical Alamouti's scheme for a $2 \times 1$ MISO system and the blind case of RIS-AP in \cite{Transmission_conference}. We observe that a second order transmit diversity is achieved using a single RF signal generator thanks to the RIS. Furthermore, the influence of the number of RIS elements on the BER performance is shown to be significant, where a 10 dB gain is achieved with $64$ RIS elements. As it can be verified from (1), the increase of the number of RIS elements enhances the BER performance linearly, where a $3$ dB gain is achieved by doubling $N$. Hence, the RIS-assisted Alamouti's scheme outperforms the classical Alamouti's scheme in terms of the BER performance and the required RF resources on the transmitter side. Also, compared to the blind scheme of RIS-AP in \cite{Transmission_conference}, we see that the proposed scheme clearly shines out by a  diversity of order two due to the orthogonality of Alamouti's transmission matrix.

For the RIS-assisted and IM-based VBLAST scheme, we considered the existence and absence of a LOS component for the S-RIS, and RIS-D channels. Therefore, the simulations are performed with $K=5, $ and $-\infty$ dB, where $K$ values can change dramatically within the same indoor environment depending on the location of the transmitter and receiver \cite{RicianIndoor}.

In Figs. 5 and 6, we compare RIS-assisted and IM-based VBLAST with classical VBLAST for $2\times2$ MIMO setup. Algorithms 1 and 2 correspond to the optimal and sub-optimal detectors for the transmit-receive antenna indices and they are denoted in Figs. 5 and 6 by ``opt." and ``sub-opt.", respectively. Here, the spectral efficiency of classical VBLAST is $2$ b/s/Hz, while it is $3$ and $4$ b/s/Hz for the RIS-assisted and IM-based VBLAST, in partial and full-IM modes, respectively, with BPSK. Thus, the spectral efficiency for the RIS-assisted IM-based VBLAST is significantly boosted compared to the classical VBLAST due to the extra IM bits conveyed by our antenna pair selection methodology.

In Fig. 5 we compare the BER performance of RIS-assisted and IM-based VBLAST scheme, under the Rayleigh fading assumption for the S-RIS-D transmission path, with classical VBLAST scheme. We observe that the proposed scheme operated in full-IM mode provides an improved BER at low to mid SNR values, while saturating to the BER performance of the classical VBLAST at high SNR. Nevertheless, compared to the classical VBLAST scheme, the spectral efficiency is doubled for the proposed scheme. In addition, the sub-optimal detector represented by Algorithm 2 is shown to achieve the same performance as for the optimal one, which is represented by Algorithm 1. This means that the detection process of the antenna indices can be simpler in terms of the complexity level. It can be also seen that the performance of the continuous phases adjustment can be almost achieved with only four discrete phase shifts \cite{BeamformingPhs}, \cite{HowmanyPhs}. In this way, the RIS design can be simplified to control the phase shift of each element using  2-bits only.   
\begin{figure}[t] 
	\begin{center}
		\includegraphics[width=3.4in,height=2.7in]{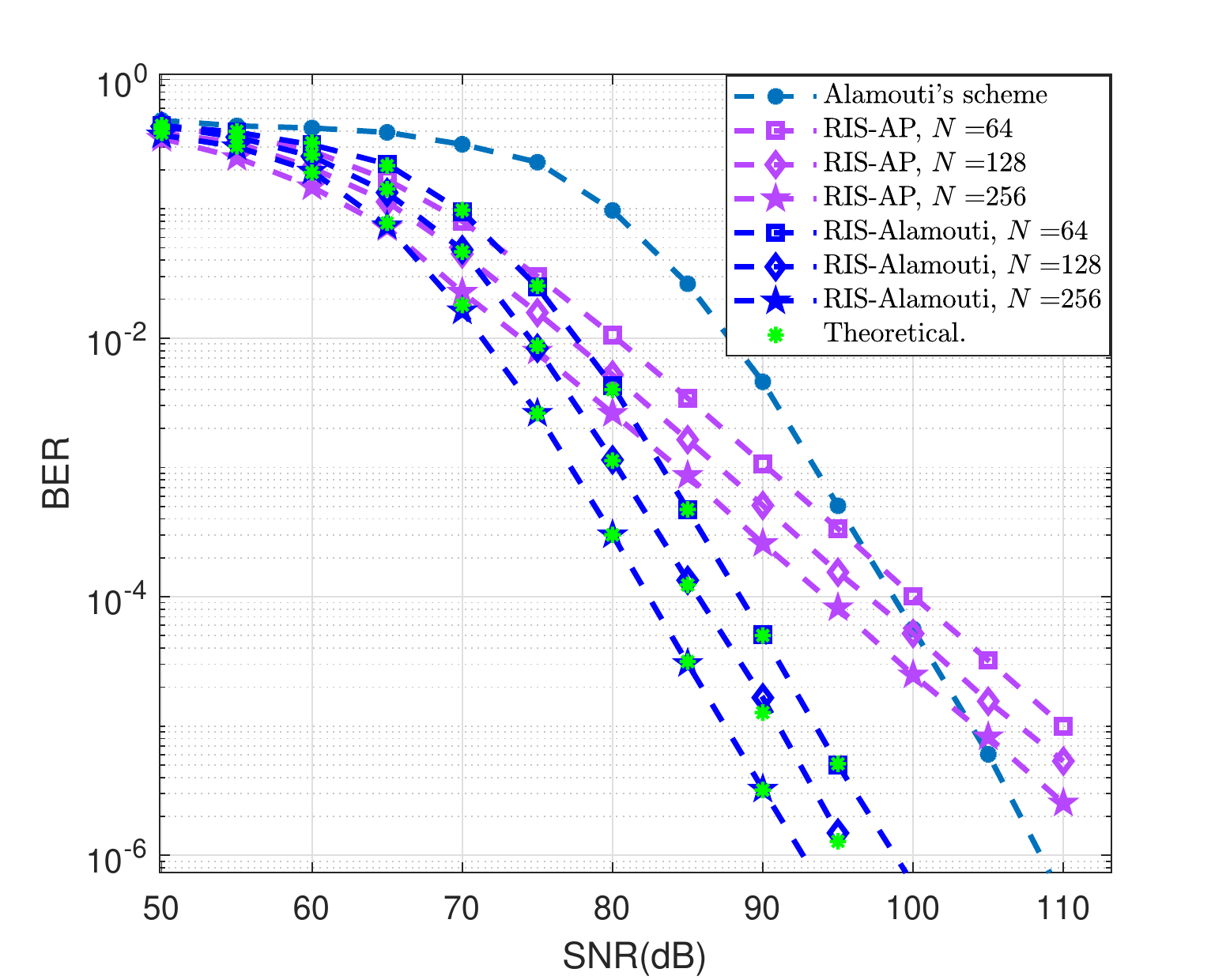}
		\caption{BER performance of RIS-assisted Alamouti’s scheme
			versus $2\times1$ classical Alamouti’s scheme and RIS-AP (Blind)
			scheme, with different N values and BPSK.}
	\end{center}
	\vspace{-0.5cm}
\end{figure}

Furthermore, the partial-IM and enhancing modes show a significant BER performance gain of $16$ and $28$ dB, respectively, compared to the classical VBLAST scheme. It is worth noting that, the partial-IM scheme increases the spectral efficiency by $50\%$.

In Fig. 6, the BER performance curves are shown for the RIS-assisted and IM-based VBLAST scheme considering the existence of LOS component ($K=5$ dB) between S and the RIS, and between the RIS and D. Compared to the classical VBLAST scheme, the impact of the LOS component on the RIS performance can be clearly seen for the proposed scheme. The full-IM and partial-IM modes need an SNR increase of $5$-$10$ dB in order to increase the spectral efficiency by $2$ and $1.5$ fold, respectively. This can be explained by the lack of diversity due to the LOS component in the S-RIS and RIS-D channels. Therefore, the receiver makes more errors while detecting the targeted antenna indices, $l^*$ and $m^*$.  Since the construction of $\hat{\mathbf{V}}$ depends on $l^*$ and $m^*$, the erroneous detection of them reflects on the detection of the $M$-QAM symbols. Furthermore, it can be noted that the sub-optimal (greedy) detector provides an error floor even with an RIS of $512$ elements. This is, also, due to the LOS component, which makes the difference in the instantaneous energy, received by all the receiving antennas, trivial.  Nevertheless, the enhancing mode still provides a BER performance gain of $4$ dB compared to the classical VBLAST scheme. 

 Comparing the three operating modes, Fig. 5 and 6 show that the enhancing mode, where there is no IM, achieves the best BER performance. This is due to the fact that, in enhancing mode, the amplification is directed towards a fixed and predefined antenna pair. Furthermore, the proposed scheme has the flexibility to operate in one of these three modes according to the $K$ value, hence, increase the spectral efficiency and/or enhance the BER performance accordingly.

 Finally, it is worth noting that, from (22) and (23), comparing the S-RIS-D and S-D communication links, we obtain $7.86$ dB additional path loss, under the given separation distances, for RIS-AP and RIS-assisted Alamouti schemes compared to the classical Alamouti's scheme. In the same way, for RIS-assisted and IM-based VBLAST scheme, where there are two communication links, there is an additional path loss of $12.29$ dB  for the S-RIS-D compared to the S-D communication link. This shows that the transmission over the RIS has considerably higher path loss compared to the direct transmission without RIS. Nevertheless, the additional path loss can be compensated by choosing the proper RIS size \cite{Ellingson}, \cite{ExperimentRIS}, as we showed in our proposed  schemes.
 \begin{figure}[t] 
 	\begin{center}
 		\includegraphics[width=3.25in,height=2.55in]{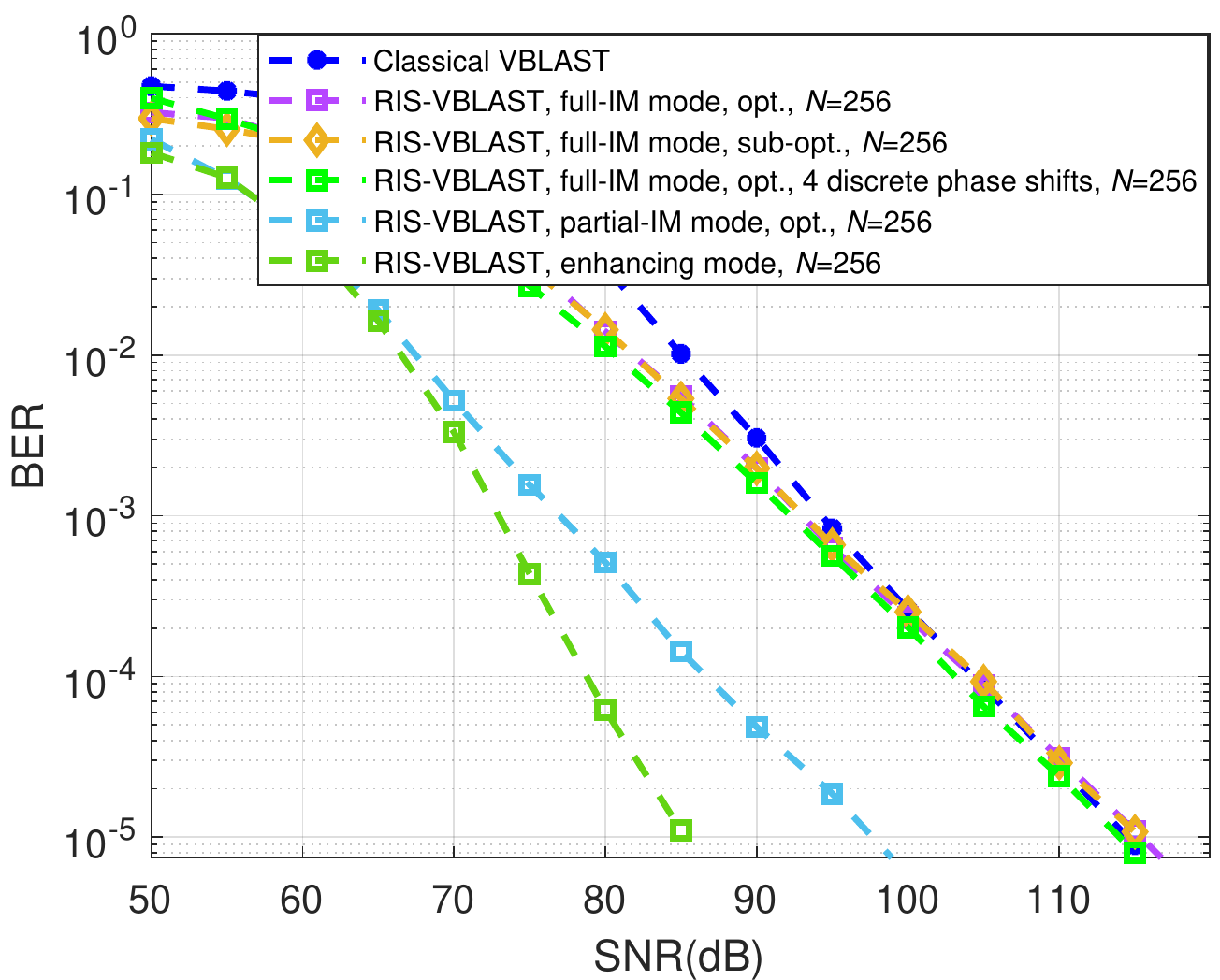}
 		\caption{BER performance of RIS-assisted and IM-based VBLAST scheme versus classical VBLAST with $2\times 2$ MIMO setup, BPSK, and $K=-\infty$ dB.}
 	\end{center}
 	\vspace{-0.5cm}
 \end{figure}
\section{Conclusion}
\vspace{-0.01cm}
 \begin{figure}[t] 
	\begin{center}
		\includegraphics[width=3.2in,height=2.5in]{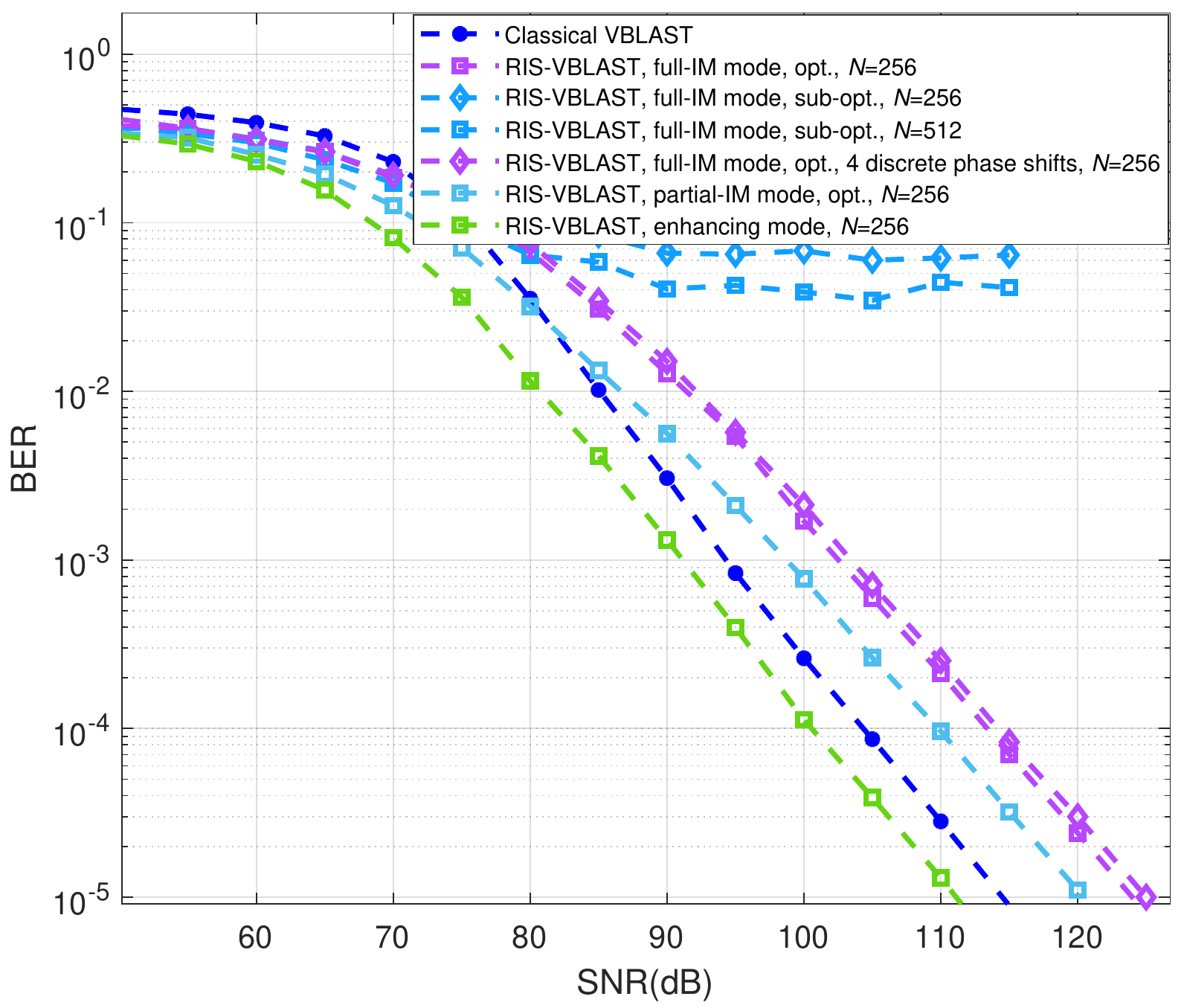}
		\caption{BER performance of RIS-assisted and IM-based VBLAST scheme versus classical VBLAST with $2\times 2$ MIMO setup, BPSK, and $K=5$ dB.}
	\end{center}
	\vspace{-0.5cm}
\end{figure}
In this paper, we have proposed novel designs for MIMO systems with the assistance of RISs. Although only VBLAST and Alamouti's schemes have been considered in this study, our concepts can be applied for other MIMO schemes as well. Applying the considered concept to space-time codes in large scale MIMO setups with a large number of antennas may show the remarkable advantage for this scheme by utilizing a single RF signal generator instead of multiple RF chains. Furthermore, RIS-assisted Alamouti's scheme is capable of achieving an $N$ times SNR enhancement in addition to a transmit diversity order of two and the RIS-assisted IM-based VBLAST scheme is able to provide a significant BER performance gain in addition to the noticeable increasing in the spectral efficiency by the smart methodology of channel phases elimination. Considering the simplicity of implementation and deployment, both schemes do not require a significant reconfiguration for the existing MIMO setups, particularly in their receiver architectures, which makes them practical and feasible alternatives for future wireless systems. The design of a practical phase shift model, that captures the phase-dependent amplitude variation, for the proposed schemes appears as interesting problem which we will consider in our future research.

\bibliographystyle{IEEEtran}
\bibliography{IEEEabrv,bibliography}

%
\vspace{4cm}
\begin{IEEEbiography}[{\includegraphics[width=1in,height=1.25in,clip,keepaspectratio]{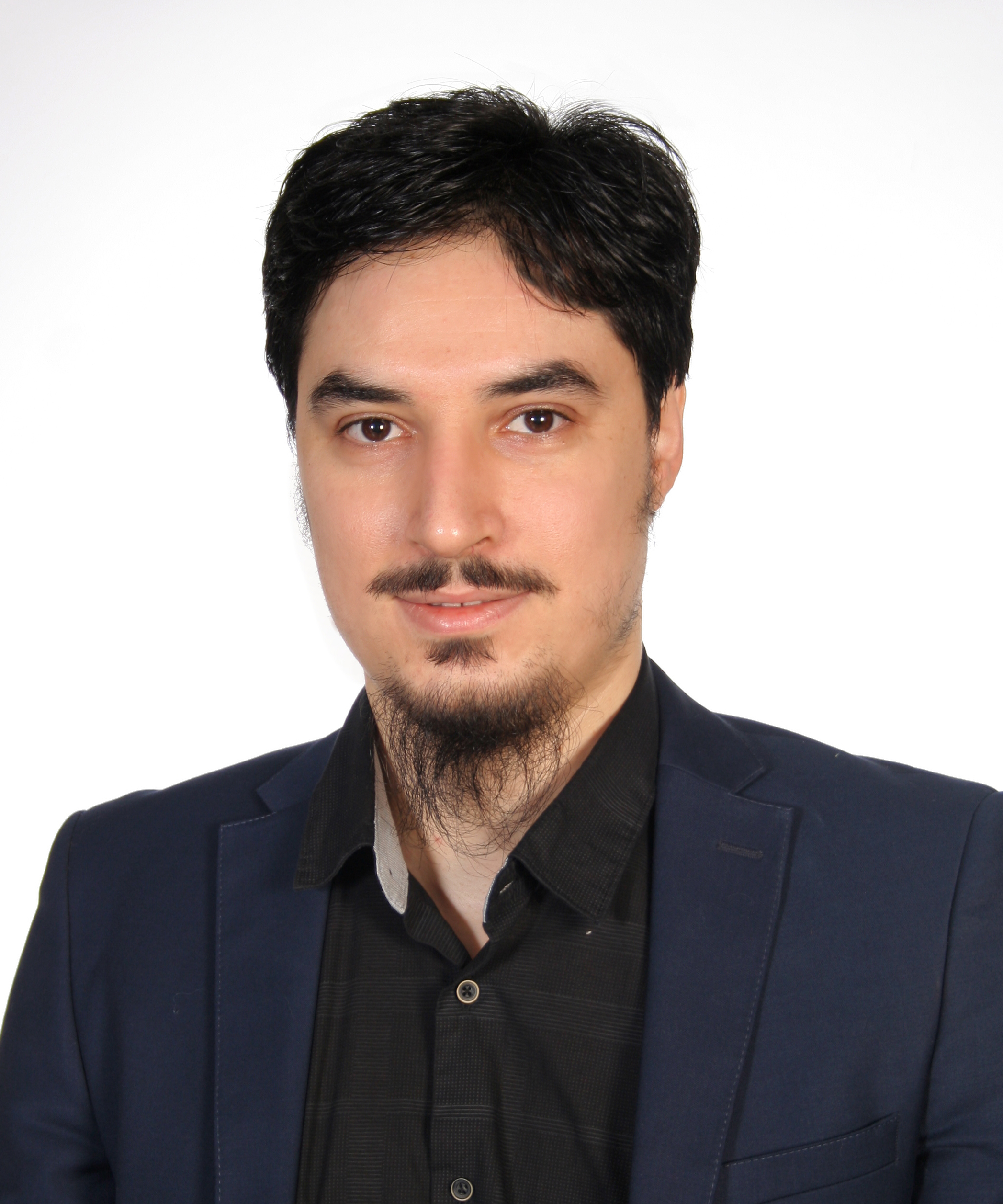}}]{Aymen Khaleel {\normalfont received the B.Sc. degree from the University of Anbar, Al Anbar, Iraq, in 2013, and the M.Sc. degree from Turkish Aeronautical Association University, Ankara, Turkey, in 2017. He is currently pursuing his Ph.D. in Electrical and Electronics Engineering at Ko\c{c} University, Istanbul, Turkey, where he is currently a Teaching Assistant. His research interests include MIMO systems, index modulation, intelligent surfaces-based systems. He serves as a Reviewer for \textit{IEEE communications letters}.}}
\end{IEEEbiography}
\begin{IEEEbiography}[{\includegraphics[width=1in,height=1.25in,clip,keepaspectratio]{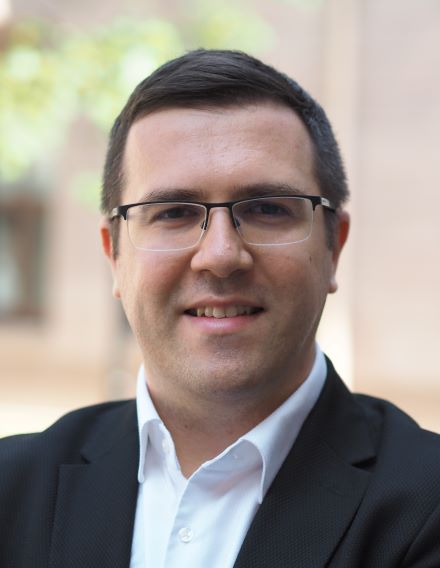}}]{Ertugrul Basar {\normalfont (S'09-M'13-SM'16) received the B.S. degree (Hons.) from Istanbul University, Turkey, in 2007, and the M.S. and Ph.D. degrees from Istanbul Technical University, Turkey, in 2009 and 2013, respectively. He is currently an Associate Professor with the Department of Electrical and Electronics Engineering, Koç University, Istanbul, Turkey and the director of Communications Research and Innovation Laboratory (CoreLab). His primary research interests include MIMO systems, index modulation, intelligent surfaces, waveform design, visible light communications, and signal processing for communications.\\ \indent Recent recognition of his research includes the IEEE Communications Society Best Young Researcher Award for the Europe, Middle East, and Africa Region in 2020, Science Academy (Turkey) Young Scientists (BAGEP) Award in 2018, Turkish Academy of Sciences Outstanding Young Scientist (TUBA-GEBIP) Award in 2017, and the first-ever IEEE Turkey Research Encouragement Award in 2017.\\
\indent Dr. Basar currently serves as a Senior Editor of the  \textit{IEEE Communications Letters} and the Editor of the \textit{IEEE Transactions on Communications}, and \textit{Physical Communication (Elsevier)}, and \textit{Frontiers in Communications and Networks}.}}

\end{IEEEbiography}







\end{document}